\documentclass[aps,11pt]{revtex4}
\usepackage{array}[=2016-10-06]
\usepackage{dcolumn}
\usepackage{graphicx}
\usepackage{amsmath}
\usepackage{amsfonts}
\usepackage{amssymb}
\usepackage{psfrag}
\usepackage{wrapfig}
\usepackage{subfigure}
\usepackage{makeidx}
\usepackage{bm}
\usepackage{epsf}
\usepackage{hyperref}
\usepackage{color}
\usepackage{cases}
\usepackage{multirow}
\usepackage{booktabs} 
\usepackage{makecell} 
\usepackage{float}
 
\makeatletter
\newcommand{\eq}[1]{Eq.(\ref{#1})}

\newcommand{\Rmnum}[1]{\uppercase\expandafter{\romannumeral #1}}
\usepackage{amsmath} 
\begin{document} 

\title{Axial gravitational quasinormal modes of magnetically charged black holes}

\author{De-Cheng Zou$^{1}$\footnote{dczou@jxnu.edu.cn}, Xufen Zhang$^{2}$\footnote{xfzhangyzu@126.com}, Chao-Ming Zhang$^{3}$\footnote{zcm843395448@163.com}, Ming Zhang$^{4,5}$\footnote{ mingzhang0807@126.com}\\ and Rui-Hong Yue$^{2}$\footnote{rhyue@yzu.edu.cn}}

\address{
$^{1}$College of Physics and Communication Electronics, Jiangxi Normal University, Nanchang 330022, China \\
$^{2}$Center for Gravitation and Cosmology, College of Physical Science and Technology, Yangzhou University, Yangzhou 225009, China \\
$^{3}$Center for Relativistic Astrophysics and High Energy Physics, Nanchang University, Nanchang, 330031, China\\
$^{4}$Faculty of Science, Xihang University, Xi'an 710077 China \\
$^{5}$National Joint Engineering Research Center for Special Pump System Technology, Xihang University, Xi’an 710077, China}


\date{\today}

\begin{abstract}
\indent

In this paper, we consider axial perturbations on the magnetically charged string-inspired Euler-Heisenberg black hole. Since axial metric perturbation decouples from axial electromagnetic perturbation, here we mainly focus on axial gravitational perturbation. By using WKB approximation and AIM methods, we make a detailed analysis of the gravitational quasinormal frequencies by varying the characteristic parameters of gravitational perturbation and black holes. It is found these results obtained through the AIM method are in good agreement with those of obtained by WKB method, including these results extracted from the time-domain profiles. The greybody factor is calculated by WKB method. The effects of these parameters $Q_m$, $\epsilon$ and multipole number $l$  on the greybody factor are also studied. 

\end{abstract}


\maketitle

\section{Introduction}
\label{intro}

The Euler–Heisenberg (EH) Lagrangian, first derived in 1936 \cite{Heisenberg:1936nmg}, serves as a nonlinear extension of quantum electrodynamics (QED) and provides a classical description that goes beyond Maxwell’s theory in strong-field regimes where vacuum polarization effects become significant. By modeling the vacuum as a polarizable medium, the EH framework incorporates effective polarization and magnetization responses arising from virtual charge fluctuations associated with real charges and currents \cite{Obukhov:2002xa}. This theoretical formalism not only refines classical electrodynamics in extreme electromagnetic environments but also forms a cornerstone for studying nonlinear phenomena in astrophysics and cosmology.

Considering these unique features, the first black hole solution in Einstein–Hilbert (EH) gravity—an anisotropic, magnetically charged configuration analogous to the Reissner–Nordström metric but generalized to include dyon-type degrees of freedom—was obtained in 1956 \cite{Yajima:2000kw}. This framework has been also extended to encompass electrically charged solutions \cite{Yajima:2000kw,Ruffini:2013hia}, rotating black hole configurations \cite{Breton:2019arv,Amaro:2022yew}, and formulations within various modified theories of gravity \cite{Guerrero:2020uhn,Nashed:2021ctg}. More recently, motivated by developments in string theory and Lovelock gravity, Ref.~\cite{Bakopoulos:2024hah} introduced a novel coupling between the dilaton field and EH electrodynamics, extending the Einstein–Maxwell–dilaton framework.
Further studies have investigated various physical phenomena in such spacetimes, including particle dynamics, gravitational lensing effects \cite{Yasir:2025npe,Vachher:2024fxs}, and the shadow characteristics of magnetically charged black holes \cite{Xu:2024gjs}. Additionally, Jiang et al. \cite{Jiang:2024njc} examined the properties of geometrically thin, optically thick accretion disks in these backgrounds, offering insights into observational signatures of such compact objects.

The recent detection of gravitational waves has opened up new avenues for probing strong gravitational fields in the vicinity of black holes. In particular, during the merger of binary compact objects, the ringdown phase of the emitted gravitational waves can be interpreted as the response of the remnant black hole to perturbations imposed during the coalescence process. At this stage, the final Kerr black hole can be effectively described as being in a perturbed state, with the emitted gravitational radiation characterized by distinct decay timescales. These features are well modeled by the quasinormal modes (QNMs) of the black hole \cite{Berti:2007,Nollert:1999}, which are unique oscillation frequencies determined solely by the black hole’s mass and spin. Furthermore, QNMs provide a powerful tool for testing the validity of the no-hair conjecture, as deviations from the expected frequency spectrum could indicate violations of the Kerr metric or the presence of exotic compact objects \cite{Berti:2006,Berti:2007b,Isi:2019}. In the context of modified theories of gravity \cite{Blazquez-Salcedo:2016}-\cite{Zhao:2022gxl}, QNMs serve as critical probes for constraining alternative gravitational models. The stability of QNMs under parametric perturbations and deformations of the background spacetime plays a key role in such investigations. However, recent studies have also shown that certain QNM spectra may exhibit instability under small perturbations to the effective potential \cite{Jaramillo:2021,Cheung:2022}. Moreover, the behavior of QNMs under various types of perturbations offers valuable insights into the stability properties of the underlying spacetime geometry \cite{Ishibashi:2003}-\cite{Yan:2020nvk}.

Motivated by these considerations, this paper focuses on the QNMs of the magnetically charged black hole. Notice that
Cho et al. \cite{Cho:2009cj,Cho:2011sf} introduced the asymptotic iteration method (AIM), a systematic iterative technique for deriving accurate approximations of QNMs frequencies. In this work, we employ AIM to numerically solve the perturbation equation. Additionally, we apply the WKB approximation \cite{Iyer:1986np,Iyer:1986nq, Konoplya:2003ii}, a well-established method, to estimate the QNMs frequencies of the black holes. By comparing the results from both approaches, we aim to cross-validate and reinforce our findings.
Another crucial aspect of black hole perturbation theory is the greybody factor \cite{Konoplya:2019ppy,Konoplya:2011qq,Konoplya:2019hlu,Gogoi:2023fow},which characterizes how a black hole's gravitational potential modifies the spectrum of emitted radiation. In gravitational wave astronomy, greybody factors and QNMs form a complementary framework for analyzing compact binary mergers: QNMs determine the ringdown phase through their characteristic frequencies and damping times, while greybody factors quantify the angular-dependent transmission probability of gravitational waves throughout the entire inspiral-merger-ringdown process \cite{Oshita:2023cjz}. Although the background geometry remains spherically symmetric, axial perturbations are particularly suitable for analyzing the dynamics of magnetically charged black holes in the string-inspired Euler-Heisenberg framework. Due to the monopolar structure of the gauge potential, the axial sector exhibits a decoupled evolution governed by a single master equation. This simplification facilitates clearer computation of QNMs and greybody factors and has been widely adopted in related studies of nonlinear electrodynamics.

In light of these, we plan to study the QNMs and greybody factor of axial perturbations on the magnetically charged black hole in string-inspired Euler–Heisenberg theory. This paper is constructed as follows. In Section \ref{sec2}, we review the black hole solution briefly and derive the master equation for axial perturbations. In Section \ref{sec3}, we solve the quasinormal frequencies with AIM and WKB methods, and study the effects of the black hole parameters on the QNMs.  In Section \ref{sec4}, we
investigate the time evolution profiles of the perturbation in Section \ref{sec3}. In Section \ref{sec5}, we calculate the greybody factor with the WKB method. The conclusions and discussions are given in Section \ref{sec6}.

\section{Master equation for black hole perturbations}
\label{sec2}

\subsection{Background and black hole solution}

Motivated by string theory and Lovelock gravity, Bakopoulos et al. \cite{Bakopoulos:2024hah} recently introduced an extension of the Einstein--Maxwell--dilaton theory that incorporates a nonlinear Euler--Heisenberg term coupled to the dilaton field. The action describing this model is given by \cite{Bakopoulos:2024hah}:
\begin{eqnarray}
S=\frac{1}{16\pi}\int  d^4x\sqrt{-g}\left(R-2\nabla^\mu\phi\nabla_\mu\phi-{\cal L}(\phi,F)\right),\label{action}
\end{eqnarray}
where $R$ denotes the scalar curvature, $\phi$ is the scalar field and ${\cal L}(\phi,F)$ represents the Lagrangian density that governs the interaction between the dilaton and the nonlinear electromagnetic field. This Lagrangian is explicitly defined as
\begin{eqnarray}
{\cal L}(\phi,F)=e^{-2\phi}F^2+f(\phi)\left(2\alpha F^\mu_{~\nu} F^\nu_{~\rho} F^\rho_{~\delta} F^\delta_{~\mu}-\beta F^4\right).
\end{eqnarray}
Here $f(\phi)$ is a coupling function, $F^2=F_{\mu\nu}F^{\mu\nu}$ and $F^4=F_{\mu\nu}F^{\mu\nu}F_{\rho\delta}F^{\rho\delta}$, where $F_{\mu\nu}$ stands for the usual field strength $F_{\mu\nu}=\partial_\mu A_{\nu}-\partial_{\nu}A_{\mu}$. 
In the case where $\alpha=\beta=0$, the theory reduces to the standard Einstein--Maxwell--dilaton model.

Varying the action \eqref{action} with respect to $g_{\mu\nu}$, $\phi$, and $A_\mu$, we can obtain the three field equations
\begin{eqnarray}
&&E^{(g)}_{\mu\nu}=R_{\mu\nu}-\frac{1}{2}R g_{\mu\nu}-2{\partial _\mu}\phi{\partial_ \nu}\phi+g_{\mu\nu}{\partial }^\mu\phi {\partial }_\mu \phi-T_{\mu\nu}=0,\label{eqG}\\
&&E^{(\phi)}=\Box\phi +\frac{1}{2}e^{-2\phi}F^2-\frac{df(\phi)}{d\phi}\left(\frac{\alpha}{2} F^\mu_{~\nu} F^\nu_{~\gamma} F^\gamma_{~\delta} F^\delta_{~\mu}-\frac{\beta}{4} F^4\right)=0,\label{eqKG}\\
&&E^{(A)}_{\nu}={\partial^ \mu}\Big[\sqrt{-g}\left(4F_{\mu\nu}(2\beta f(\phi)F^2-e^{-2\phi})-16\alpha F^{\mu\kappa} F^\kappa_{~\lambda} F_{\nu}^{\lambda}\right)\Big]=0,\label{eqMaxwell}
\end{eqnarray}
where $T_{\mu\nu}$ is energy-momentum tensor with
\begin{eqnarray}
T_{\mu\nu}&=&2e^{-2\phi}(F^{\alpha}_\mu F_{\nu \alpha}-\frac{1}{4}g_{\mu\nu }F^2)\nonumber\\
&&+f(\phi)\left ( 8\alpha F^\alpha _\mu F^\beta_\nu F^\eta_\alpha F_{\beta\eta} -\alpha g_{\mu\nu} F^\alpha_\beta F^\beta _\gamma F^\gamma _\delta F^\gamma_\alpha -4\beta F^\xi_\mu F_{\nu\xi}F^2+\frac{1}{2}g_{\mu\nu}\beta F^4\right).
\end{eqnarray}

With regard to the coupling function  $f(\phi)$, Bakopoulos et al.\cite{Bakopoulos:2024hah} adopt
\begin{eqnarray}
 f(\phi)=-\Big[3\text{cosh}(2\phi)+2\Big]\equiv-\frac{1}{2}\left(3e^{-2\phi}+3e^{2\phi}+4\right),\label{phi}
\end{eqnarray}
and then obtain an exact analytic solution in the presence of magnetic charge and scalar hair. This hyperbolic structure reflects a symmetric dilaton coupling often encountered in string-theoretic effective actions, ensuring that both $\phi\rightarrow\infty$ and $\phi\rightarrow-\infty$ regimes are covered in a regular manner. Moreover, this form guarantees a nontrivial contribution from higher-order nonlinear electromagnetic terms and stabilizes the potential structure needed for horizon formation.
Defining $\epsilon=\alpha-\beta$, we  obtain
the magnetically charged black
hole solution \cite{Bakopoulos:2024hah}
\begin{eqnarray}
 ds^2&=&-H(r) \,dt^2 + \frac{1}{H(r)} \, dr^2 + R(r)^2 \,( d\theta^2 + \sin^2 \theta \, d\varphi^2),\label{oldmetric}\\
H(r)&=&1-\frac{2 M}{r}-\frac{2\epsilon Q_m^4}{r^3(r-Q_m^2/M)^3}, \quad
R(r)=r\left(r-\frac{Q_m^2}{M}\right),\nonumber\\
\phi (r)&=&-\frac{1}{2}\ln \left(1-\frac{Q_m^2}{M r}\right), \quad A_\mu=(0, 0, 0, Q_m\cos\theta),\label{Qm}\label{fRsolution}
\end{eqnarray}
where $M$ and $Q_m$ are the mass and magnetic charge of this black hole, respectively.  

In the limit $\epsilon=0$,  the solution \eqref{fRsolution} reduces to the GMGHS or GHS black holes \cite{Gibbons:1987ps,Garfinkle:1990qj}, which have been extensively studied \cite{Ferrari:2000ep,Chen:2004zr,Shu:2004fj,Karimov:2018whx}. Moreover, this solution \eqref{fRsolution} describes a black hole with a single horizon when $\epsilon=1$, while for $\epsilon=-1$, the black hole horizons can range from two to none. These imply the magnetically charged black hole possesses different horizon structures. 

In order to consider the QNMs and greybody factor of axial  perturbation on the magnetically charged black hole in string-inspired Euler–Heisenberg theory, we rewrite the metric \eqref{oldmetric} and solution \eqref{fRsolution} into new forms
\begin{eqnarray}
  ds^2 & =& -A(r)dt^2 + \frac{1}{B(r)} dr^2 + r^2 (d\theta^2 + \sin^2 \theta d\varphi^2),\label{metric}\\
A(r)&=&1-\frac{4 M^2}{Q_m^2+\sqrt{Q_m^4+4 M^2 r^2}}-\frac{2 \epsilon Q_m^4}{r^6},\nonumber\\
B(r)&=&1
- \frac{Q_{m}^{4} + 4M^{2}r^{2}}{r^2(Q_{m}^{2} + \sqrt{Q_{m}^{4} + 4M^{2}r^{2}})} +\frac{Q_m^4}{4M^2r^2}
- \frac{\epsilon Q_{m}^{4}(Q_{m}^{4} + 4M^{2}r^{2})}{2M^2r^{8}},\nonumber\\
\phi (r)&=&-\frac{1}{2}\ln \left(\frac{\sqrt{Q_m^4+4 M^2 r^2}-Q_m^2}{\sqrt{Q_m^4+4 M^2 r^2}+Q_m^2}\right)\label{solution}.
\end{eqnarray}

\subsection{Axial perturbations}

Here we focus on the axial perturbation for the magnetically charged black holes. To do so, we assume
\begin{eqnarray}
  g_{\mu\nu}=\bar{g}_{\mu\nu}+\delta{g}_{\mu\nu},\quad  A_{\mu}=\bar{A}_{\mu}+\delta{A}_{\mu},\label{pertmetric}
\end{eqnarray}
where $\bar{g}_{\mu\nu}$ and $\bar{A}_{\mu}$ represent the background metric and electromagnetic field, and $\delta{g}_{\mu\nu}$ and $\delta{A}_{\mu}$ denote the corresponding perturbations. 

Under the Regge–Wheeler gauge \cite{Regge:1957td}, we expand the metric perturbation using tensor spherical harmonics. The axial gravitational field perturbation involves two modes $h_0(r)$ and $h_1(r)$, and the perturbed metric is expressed as
\begin{eqnarray}
\delta{g}_{\mu\nu}=
\sum_{l,m} e^{-i\omega t}\begin{bmatrix}
0 & 0 &0 & h_0(r) \\
0 & 0 &0 & h_1(r) \\
0 & 0 & 0 & 0 \\
h_0(r) & h_1(r) & 0 & 0
\end{bmatrix}\sin\theta\partial_{\theta}Y_{lm}, \label{eqodd}
\end{eqnarray}
where the spherical harmonics can be replaced by Legendre
polynomials by setting the azimuthal number $m=0$ without
loss of generality, i.e., $Y_{lm}|_{m=0}=\sqrt{\frac{2l+1}{4\pi}}P_{l}(\cos\theta)$, because
the background metric is spherically symmetric.

Following
Refs.\cite{Zerilli:1970se,Brito:2018hjh}, axial vector perturbation is given by
\begin{eqnarray}\label{vectorpert}
\delta{A}_{\mu}=\sum_{l,m} e^{-i\omega t}\Big[0,0,-u_3(r)\frac{\partial_{\varphi}Y_{lm}}{\sin\theta},u_3(r)\sin\theta\partial_{\theta}Y_{lm}\Big].
\end{eqnarray}
Substituting the perturbed metric and vector potential (\eqref{pertmetric}, \eqref{eqodd} and \eqref{vectorpert}) into the gravitational field equation \eqref{eqG}, the non-zero components of first-order perturbed gravitational field equation are obtained as
\begin{eqnarray}
E^{(g)}_{tt}&=&E^{(g)}_{rr}=E^{(g)}_{t\theta}= \Big[r^4-2\epsilon Q_m^2\left(4
   e^{2 \phi}+3 e^{4 \phi}+3\right)\Big]u_3,\label{maxpert1}\\
E^{(g)}_{r\theta}&=&\Big[r^4-2\epsilon Q_m^2\left(4
   e^{2 \phi}+3 e^{4 \phi}+3\right)\Big]u_3',\label{maxpert2}\\
E^{(g)}_{\theta \theta}&=&E^{(g)}_{\varphi\varphi}=\Big[r^4-6\epsilon Q_m^2\left(4
   e^{2 \phi}+3 e^{4 \phi}+3\right)\Big]u_3,\label{maxpert3}\\
E^{(g)}_{t\varphi}&=&r^2 B h_0 A'^2-r^2 A \Big[h_0A' B'+B(i\omega  h_1 A'+A' h_0'+2h_0A'')\Big]-A^2\Big[2h_0\left(r^2 {\cal \bar{L}}(\phi,F)+2r B'\right.\nonumber\\
&&\left.-2+l(l+1)+2B(1+r^2\phi'^2)\right)-r^2B'\left(i\omega h_1+ h_0'\right)-2rB\left(2i\omega h_1+i\omega r h_1'+r h_0''\right)\Big],\label{E14}\\
E^{(g)}_{r\varphi}&=&-r^2 B h_1 A'^2+2 A^2 h_1\left(r^2 {\cal \bar{L}}(\phi,F)-2+l+l^2+r B'+2r^2B\phi'^2\right)\nonumber\\ 
&&-r A \Big[4i\omega h_0-2i r\omega h_0'+h_1\left(r(2\omega^2-A' B')-2B(A'+rA'')\right)\Big],\label{E24}\\
E^{(g)}_{\theta\varphi}&=&2i\omega h_0+ A h_1 B'+B(h_1 A' + 2 A h_1'),\label{E34}
\end{eqnarray}
where ${\cal \bar{L}}(\phi,F)$ denotes the Lagrangian
density for black hole background solutions \eqref{solution}, and reads as 
\begin{eqnarray}
{\cal \bar{L}}(\phi,F)=\frac{2 Q_m^2 e^{-2\phi (r)}}{r^4}.\label{LFF}
\end{eqnarray}

From Eqs.\eqref{maxpert1}- \eqref{maxpert3}, one can easily find the perturbation function $u_3(r)=0$. Moreover, the electromagnetic perturbation function $u_3(r)$ does not appear in gravitational perturbation equations (\eqref{E14}, \eqref{E24} and \eqref{E34}). It implies that the axial metric perturbation decouples from axial electromagnetic perturbation. In fact, the phenomenon has been recovered in Ref. \cite{Daghigh:2021psm} for the magnetic black holes.
In addition, we substitute the perturbed metric and vector potential (\eqref{pertmetric}, \eqref{eqodd} and \eqref{vectorpert}) into electromagnetic field equation \eqref{eqMaxwell}, and then obtain corresponding perturbed equations. The detail has been shown in Appendix, where it also recovers this decoupling phenomenon. Then, we obtain a single schr\"odinger like equation for perturbed electromagnetic field $u_3(r)$. 

It's well known that gravitational perturbation are central to black hole physics because they directly probe spacetime geometry, align with observational priorities in gravitational wave astronomy, and reflect the black hole’s intrinsic properties. In the subsequent section, we only focus on the axial gravitational perturbation.
From Eqs.\eqref{E14}, \eqref{E24} and \eqref{E34}, it's easily verified that only two of the above three
equations are independent.  Considering the component $E_{\theta\varphi}$ \eqref{E34}, we have 
\begin{eqnarray}
h_0=\frac{i}{2\omega}\Big[B(h_1 A'+2A h_1') +A h_1 B'\Big]. \label{h0}
\end{eqnarray}
Substituting \eq{h0} into the component $E_{r\varphi}$ \eqref{E24}, we can eliminate $h_0(r)$ and $h_0'(r)$, and then obtain a single second order differential equation for $h_1(r)$.
In order to cast this master equation into the standard Schr\"odinger form, we further define the function $\Psi(r)$ with
\begin{eqnarray}
h_1(r)=C_0(r)*\Psi(r).
\end{eqnarray}

We assume the function $C_0(r)$ taking following form
\begin{eqnarray}
C_0(r)=\frac{r}{\sqrt{A(r)B(r)}},
\end{eqnarray}
and then obtain the final perturbed equation
\begin{eqnarray}
 \frac{d^2\Psi(r_*)}{d r^2_{*}}+\Big[\omega^2-V(r)\Big]\Psi(r_*)=0,\label{perteq}
\end{eqnarray}
where $r_*$ represents the tortoise coordinate with
\begin{eqnarray}
d{r_*} = \frac{1}{\sqrt{AB}}dr
\end{eqnarray}
and the effective potential $V(r)$ is
\begin{eqnarray}
V(r)& =&B (r) A''(r)+A'(r)\left(\frac{B'(r)}{2}
+ \frac{B(r)}{2r}\right)-\frac{B(r)A'(r)^2} {2A(r)}\nonumber\\
&&+A(r)\Big[\frac{B'(r)}{2r}+\frac{2B (r)}{r^2}+2B(r)\phi'(r)^2+\frac{l^2+l-2}{r^2}+{\cal \bar{L}(\phi,F)}\Big]\label{oddpotv}
\end{eqnarray}
with the background Lagrangian density ${\cal \bar{L}}(\phi,F)$, see \eq{LFF}. 
We should mention that perturbation of the energy-momentum tensor of electromagnetic part can not be ignored. The similar phenomenon also appears in Ref.\cite{Wu:2018xza}.

\begin{figure}[H]
\centering
\subfigure[$\epsilon=1$]{\label{fig11} 
\includegraphics[width=2in]{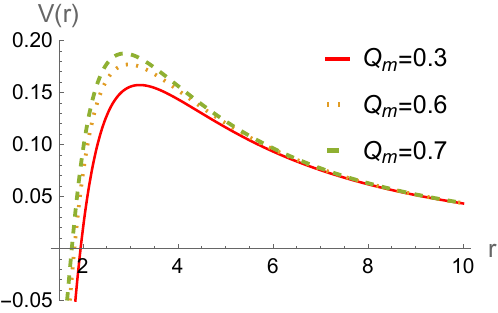}}
\hfill
\subfigure[$\epsilon=0$]{\label{fig12} 
\includegraphics[width=2in]{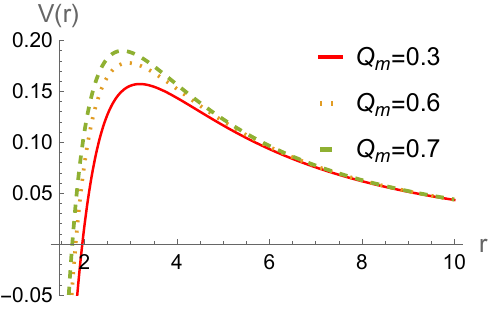}}
\hfill
\subfigure[$\epsilon=-1$]{\label{fig13} 
\includegraphics[width=2in]{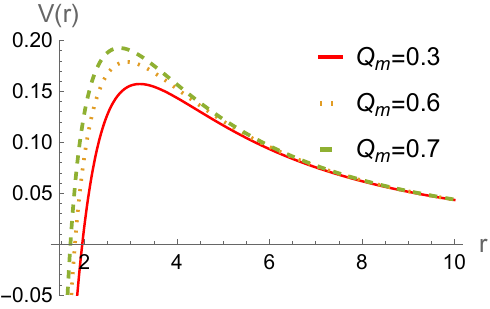}}
\caption{The effective potential $V(r)$ for the gravitational perturbation with $M=1$ and $l=2$.}\label{figpov1}
\end{figure}

\begin{figure}[H]
\centering
\subfigure[$\epsilon=1$]{\label{fig21} 
\includegraphics[width=2in]{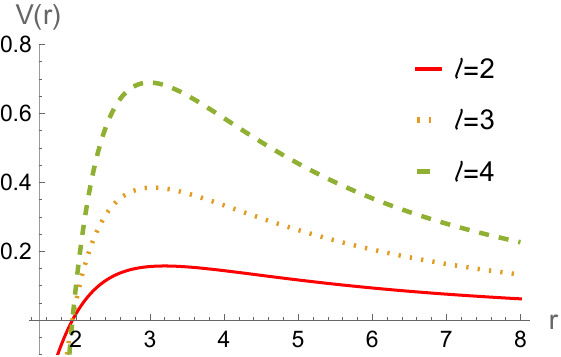}}
\hfill
\subfigure[$\epsilon=0$]{\label{fig22} 
\includegraphics[width=2in]{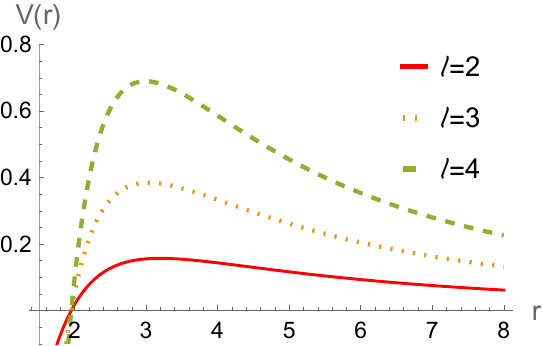}}
\hfill
\subfigure[$\epsilon=-1$]{\label{fig23} 
\includegraphics[width=2in]{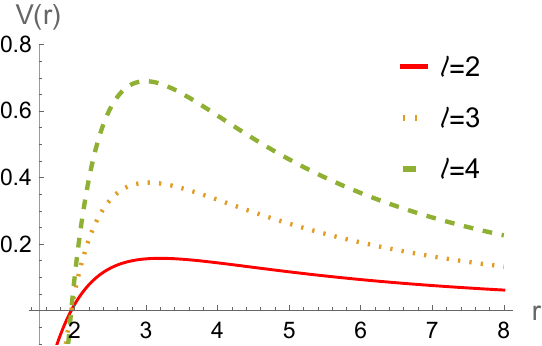}}
\caption{The effective potential $V(r)$ for the gravitational perturbation with $M=1$ and $Q_m=0.3$.}\label{figpov2}
\end{figure}

\begin{figure}[H]
\centering
\subfigure[$Q_m=0.3$]{\label{fig30} 
\includegraphics[width=2.in]{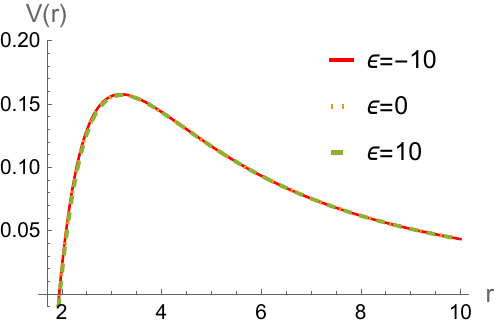}}
\hfill
\subfigure[$Q_m=0.6$]{\label{fig31} 
\includegraphics[width=2.in]{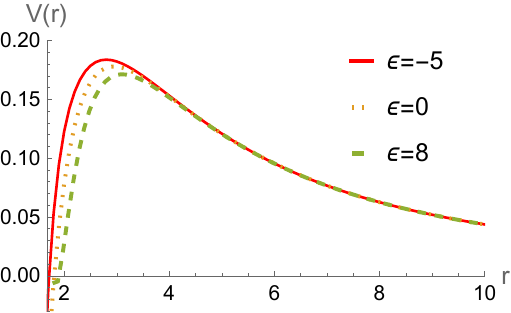}}
\hfill
\subfigure[$Q_m=0.7$]{\label{fig32} 
\includegraphics[width=2.in]{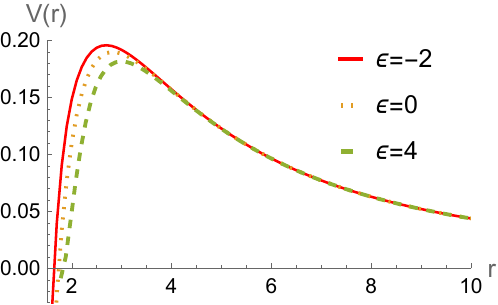}}
\hfill
\caption{The effective potential $V(r)$ for the gravitational perturbation with $M=1$ and $l=2$.}\label{figpov3}
\end{figure}

Note that the solutions \eqref{solution} and potential \eqref{oddpotv} are invariant under the following rescaling: $r/M\rightarrow r$, $Q_m /M\rightarrow Q_m$, and  $\epsilon/ M^2\rightarrow \epsilon$. 
For a better analysis of the potential function's behavior, 
We will set $M=1$ throughout the paper and leave $\epsilon$ and $Q_m$ free without loss of generality. The effective potentials $V(r)$ are plotted in Figs.\ref{figpov1}, \ref{figpov2} and \ref{figpov3}, respectively. It is seen that the height of the effective potential increases as $Q_m$ increases in Fig.\ref{figpov1}. Moreover, the potential also increases for an increase the multipole moment $l$ in Fig.\ref{figpov2}. For different $Q_m$, the effective potential exhibits distinct behaviors as the parameter $\epsilon$ varies, as shown in Fig. \ref{figpov3}. It's worthy to point out that the effective potentials are always positive, indicating that the system is stable under the axial gravitational field perturbation.

\section{Quasinormal Mode frequencies}
\label{sec3}

To accurately determine the QNMs frequencies of magnetically charged black holes, we employ two distinct computational approaches: AIM and  WKB approximation method. This dual-methodology approach enables us to perform cross-validation of our results.

\subsection{Asymptotic iteration method}

In Refs.\cite{Cho:2009cj,Cho:2011sf}, the authors applied the Asymptotic Iteration Method (AIM) to the computation of quasinormal modes (QNMs). Then, this method has been widely employed in the analysis of black hole perturbations across a variety of spacetime geometries. In the present study, we utilize AIM to numerically solve the axial gravitational perturbation equation \eqref{perteq}, enabling us to determine the QNM frequencies associated with the considered black hole background.

We can rewrite the gravitational perturbation equation \eqref{perteq} in terms of $u=1-r_+/r$ into
\begin{eqnarray}
  &&\Psi''(u)+\frac{1}{2} \left(\frac{A'(u)}{A(u)}+\frac{B'(u)}{B(u)}+\frac{4}{u-1}\right) \Psi'(u) \nonumber\\
  &&+\frac{r_+^2}{(u-1)^4A(u)B(u)}\Bigg[\omega^2+\frac{(u-1)^2}{2r_h^8}\Big(-4e^{-2\phi(u)}r_+^4(u-1)^2A(u)Q_m^2\nonumber\\
  &&+\frac{r_+^6(u-1)^2B(u){A}'(u)^2}{A(u)}-A(u)\Big(-8(\alpha-\beta)(u-1)^6(2+3\cosh(2\phi(u)))Q_m^4\nonumber\\
  &&+r_+^6\Big(2(l^2+l-2)-(u-1){B}'(u)\Big)+4r_+^6B(u)\Big(1+(u-1)^2{\phi}'(u)^2\Big)\Big)\nonumber\\
  &&-r_+^6(u-1)\Big({A}'(u)\Big(3B(u)+(u-1){B}'(u)\Big)+2(u-1)B(u){A}''(u)\Big)\Bigg]\Psi(u)=0\label{equ1}.
\end{eqnarray}
Here the range of $u$ satisfies $0\leqslant u<1$.
At the black hole horizon, the boundary conditions are pure ingoing waves
$(\Psi\sim e^{-i\omega r_*},~ r_*\to -\infty)$,
and pure outgoing waves
$(\Psi\sim e^{i\omega r_*},~ r_*\to +\infty)$,
at the spatial infinity.

To propose an ansatz for \eq{equ1}, we will examine the behavior of the function $\Psi(u)$ at horizon $(u=0)$ and at the boundary $u=1$. Near the horizon $(u=0)$, we have $A(0)\approx u A'(0)$ and $B(0)\approx u B'(0)$. Thus, \eq{equ1} reduces to
\begin{eqnarray}
  \Psi''(u)+\frac{1}{u}\Psi'(u)+\frac{r_+^2\omega^2}{u^2 A'(0) B'(0)}\Psi(u)=0.
\end{eqnarray}
Then we can obtain the solution 
\begin{eqnarray}
  \Psi(u\to 0)\sim C_1 u^{-\xi}+C_2 u^{\xi},~ \xi=\frac{ir_+\omega}{\sqrt{A'(0) B'(0)}},
\end{eqnarray}
where we have to set $C_2=0$ in order to respect the ingoing condition at the black hole horizon.

At infinity $(u=1)$, the asymptotic form of \eq{equ1} can be written as
\begin{eqnarray}
  \Psi''(u)-\frac{2}{1-u}\Psi'(u)+\frac{r_+^2\omega^2}{(1-u)^4}\Psi(u)=0,
\end{eqnarray}
which has the solution
\begin{eqnarray}
  \Psi(u\to 1)\sim D_1 e^{-\zeta}+D_2 e^{\zeta},~ \zeta=\frac{ir_+\omega}{1-u}.
\end{eqnarray}
In order to impose the outgoing boundary condition, we should set $D_1=0$.

From above solutions at horizon and infinity, we can define the general ansatz for \eq{equ1} as
\begin{eqnarray}
  \Psi(u)=u^{-\xi}e^{\zeta}\chi(u) \label{eqschi}.
\end{eqnarray}
Substituting \eq{eqschi} to \eq{equ1}, we have
\begin{eqnarray}
  \chi''=\lambda_0(u)\chi'+s_0(u)\chi \label{eqschieq},
\end{eqnarray}
where 
\begin{eqnarray}
  \lambda_0(u)=\frac{1}{2} \left(\frac{4 i r_+ \omega }{u \sqrt{A'(0)} \sqrt{B'(0)}}-\frac{A'(u)}{A(u)}-\frac{B'(u)}{B(u)}-\frac{4 \left(i r_+ \omega +u-1\right)}{(u-1)^2}\right),
\end{eqnarray}
and

\begin{eqnarray}
s_0(u)&=&\frac{1}{2}\Bigg[ \frac{2r_+\omega\Big( r_+ (u-1)^2\omega+i(u^2-1+2ir_+u\omega)\sqrt{{A}'(0)}\sqrt{{B}'(0)}\Big) }{(u-1)^2u^2{A}'(0){B}'(0)}-\frac{{A}'(u)^2}{A(u)^2}+\frac{2{A}''(u)}{A(u)}\nonumber\\
&&+ \frac{4(u-1)^2+2r_+^2\omega^2}{(u-1)^4} +4{\phi}'(u)^2+\frac{{A}'(u)}{B(u)-uB(u)}+\frac{1}{B(u)}\Big( \frac{2(l^2+l-2)}{(u-1)^2}+\frac{4e^{-2\phi(u)}Q^2_m}{r_+^2}\nonumber\\
&&-\frac{8(\alpha-\beta)(u-1)^4(2+3\cosh(2\phi(u)))Q_m^4}{r_+^6}-i r_+\omega{B}'(u)\Big( \frac{1}{(u-1)^2} -\frac{1}{u\sqrt{{A}'(0)}\sqrt{{B}'(0)}}\Big)\nonumber\\
&&+\frac{1}{A(u)}\Big( {A}'(u){B}'(u)-\frac{2r_+^2\omega^2}{(u-1)^4}\Big)\Big)+\frac{{A}'(0)}{A(u)} \Big( \frac{3u-3-i r_+\omega}{(u-1)^2}+\frac{i r_+\omega}{u\sqrt{{A}'(0)}\sqrt{{B}'(0)}}\Big)\Bigg].
\end{eqnarray}
Finally we obtain the functions $\lambda_0$ and $s_0$ for the gravitational perturbation equation \eqref{eqschieq}.

\subsection{Pad\'e averaged WKB method}

The WKB method is a well-established approach to solve black hole perturbation equation in the frequency domain \cite{Iyer:1986np}-\cite{Konoplya:2011qq}. However, the accuracy of the estimated frequencies degrades for \(n \geq l\). To ameliorate this issue, the Pad\'e approximation \cite{Matyjasek:2017psv} can be used to evaluate QNMs with higher precision. 
Within this method, the oscillation frequency $\omega$ can be determined using the following expression
\begin{eqnarray}
\omega=\sqrt{-i\Big[(n+1/2)+\sum_{k=2}^{6}\bar{\Lambda}_k\Big]\sqrt{-2V_0''}+V_0},
 \end{eqnarray}
where $n = 0, 1, 2 . . .$ stands for overtone number, $V_0 =V|_{r=r_{max}}$ and $V_0'' =\frac{d^2V}{dr^2}|_{r=r_{max}}$. The position $r_{max}$ corresponds to the location where the potential function $V(r)$ attains its highest value. 
To achieve higher precision in the calculations, correction terms $\bar{\Lambda}_k$ are introduced. The analytical expressions for these terms, together with the methodology for Pad\'e averaging, are comprehensively presented in Refs. \cite{Konoplya:2019hlu,Eniceicu:2019npi}.

The Table.\ref{Table1} shows the fundamental quasinormal frequencies $(n=0)$ of black holes with $\epsilon=0$ and $\epsilon=\pm1$ obtained through the application of the AIM and 6th-order Pad\'e averaged WKB approximation methods for different values of $Q_m$ with $M=1$ and $l=2$. Obviously, these fundamental quasinormal frequencies obtained by numerical methods agree well with each other for each branch of these black holes.

The term \(\Delta_6\) in 6th Column of Table.\ref{Table1} serves as a term to quantify the error between two adjacent order approximations, defined as \cite{Konoplya:2003ii}
\begin{eqnarray}
\Delta_6 = \frac{|\omega_7 - \omega_5|}{2},
\end{eqnarray}
where $\omega_7$ and $\omega_5$ represent the QNMs computed using the 7th and 5th-order Pad\'e-averaged WKB methods, respectively. By analyzing these errors, we provide a detailed assessment of the accuracy of our QNM estimates. 
We further consider the percentage deviation $\Delta_{AW}$ of QNMs obtained via the AIM and Pad\'e-averaged WKB methods. The relative error $\Delta_{AW}$ between two methods is defined by
\begin{eqnarray}
\Delta_{AW}=\frac{|\omega_{AIM}-\omega_{PWKB}|}{|\omega_{PWKB}|}\times 100\%.
\end{eqnarray}

\begin{table*}[h!]
\caption{Fundamental QNM frequencies for gravitational field perturbation with $M=1$.}\label{Table1}
\begin{tabular}{|m{1em}|m{1em}|m{2em}|c|c|c|c|} \hline
$\epsilon$& $l$ &  $Q_m$ &  AIM  &  Pad\'{e} averaged WKB  &  $\Delta_6$    & $\Delta_{AW}$           \\ \hline
\multirow{3}{*}{1}& \multirow{3}{*}{2}& 0.3  & $0.3812939-0.0895414i$  & $0.381247 - 0.0895259i$  & $0.0000272575$     &  $0.00412788\%$         \\ 
  &  & 0.6 &$0.4055129-0.0928762i$ & $0.405312 - 0.0929958  i$  &$0.0000877586 $  & $0.0561951\%$\\ 
  &  & 0.7 &$0.4174759-0.0960118i$  & $0.41732 - 0.0962066 i$  &$0.000171463$  &$0.0583149\%$ \\ \hline
  \multirow{3}{*}{0}&  \multirow{3}{*}{2}& 0.3  & $0.3813581-0.0894781i$  & $0.381338 - 0.0894572 i$  & $0.0000259841$     &  $0.00733116\%$         \\ 
   & & 0.6 &$0.4071892-0.0912672i$ & $0.407187 - 0.0912665   i$  &$4.24181\times10^{-6} $  & $0.0006046086\%$\\ 
  &  & 0.7 &$0.4214127-0.0922712i$  & $0.421409 - 0.0922722 i$  &$1.8842\times10^{-6}$  &$0.000862789\%$ \\\hline
   \multirow{3}{*}{-1}&  \multirow{3}{*}{2}& 0.3  & $0.3814223-0.0894146i$  & $0.381425 - 0.0893987i$  & $0.0000502269$     &  $0.00334693\%$         \\ 
   & & 0.6 &$0.4091798-0.0892181i$ & $0.40922 - 0.0893678  i$  &$0.000308716 $  & $0.0370056\%$\\ 
  &  & 0.7 &$0.4261004-0.0874892i$  & $0.426335 - 0.0878444 i$  &$0.00114384$  &$0.0978375\%$ \\\hline
 \end{tabular}
\end{table*}

\begin{figure}[H]
\centering
\subfigure[$\epsilon=1 $]{\label{figQ11} 
\includegraphics[width=2.3in]{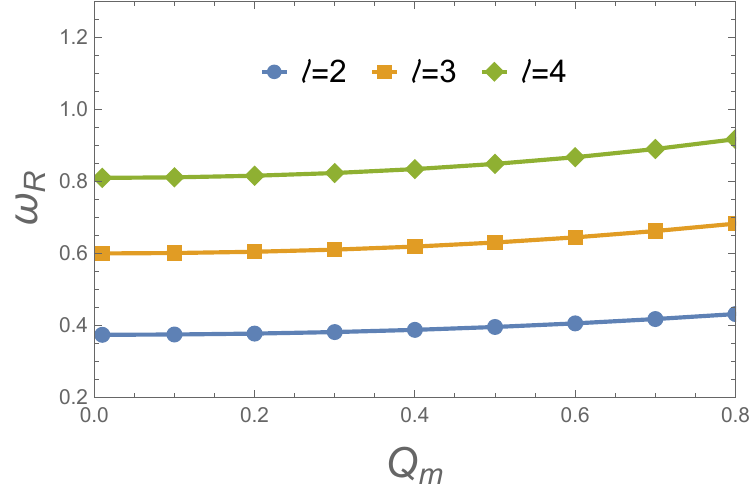}}
\subfigure[$\epsilon=1$]{\label{figQ12} 
\includegraphics[width=2.4in]{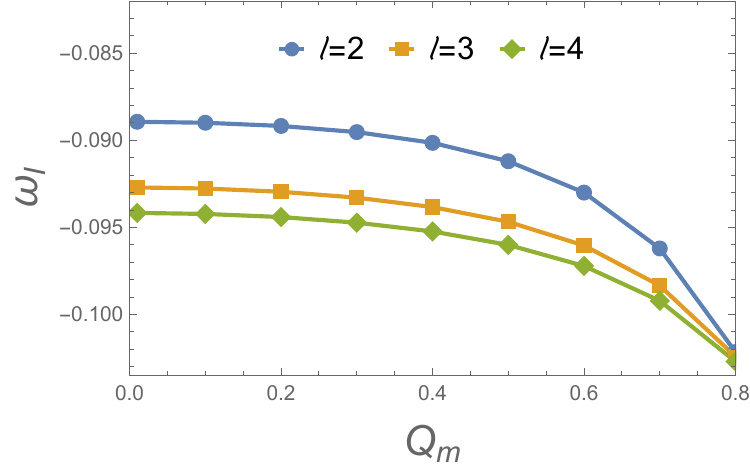}}
\subfigure[$\epsilon=0 $]{\label{figQ15} 
\includegraphics[width=2.3in]{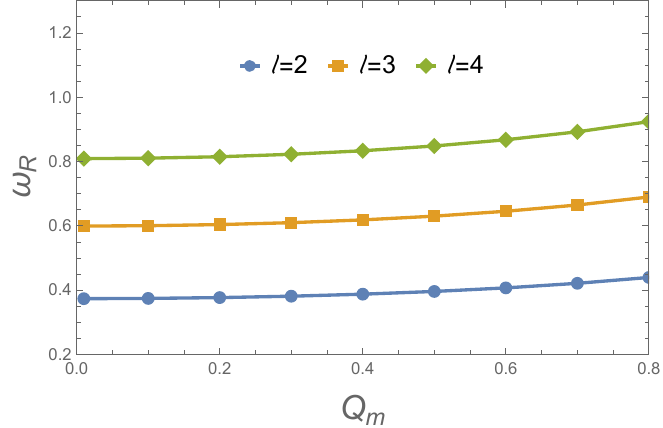}}
\subfigure[$\epsilon=0$]{\label{figQ16} 
\includegraphics[width=2.4in]{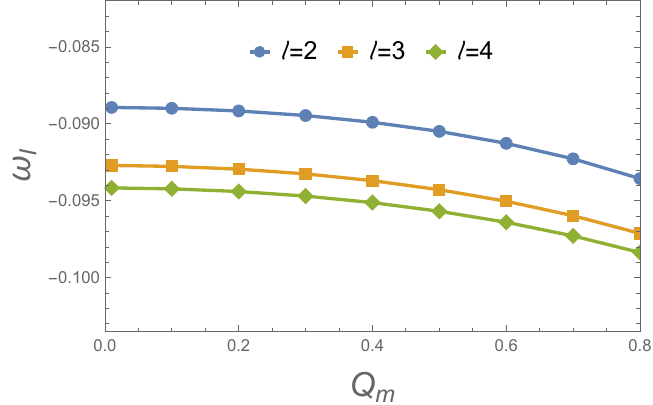}}
\subfigure[$\epsilon=-1$]{\label{figQ13} 
\includegraphics[width=2.3in]{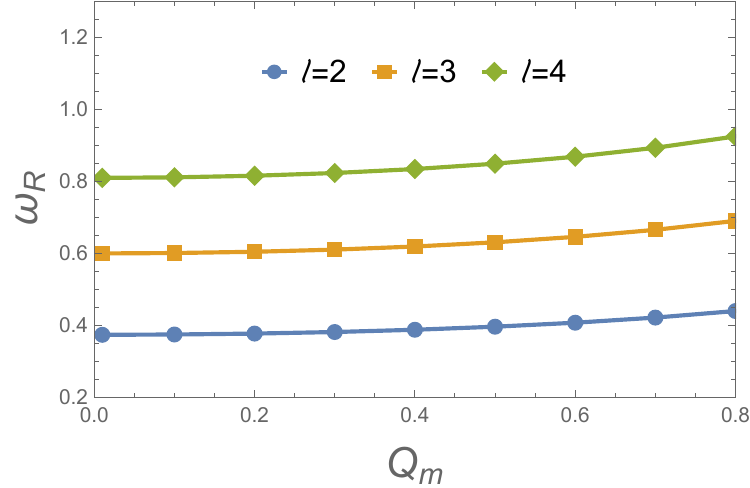}}
\subfigure[$\epsilon=-1$]{\label{figQ14} 
\includegraphics[width=2.4in]{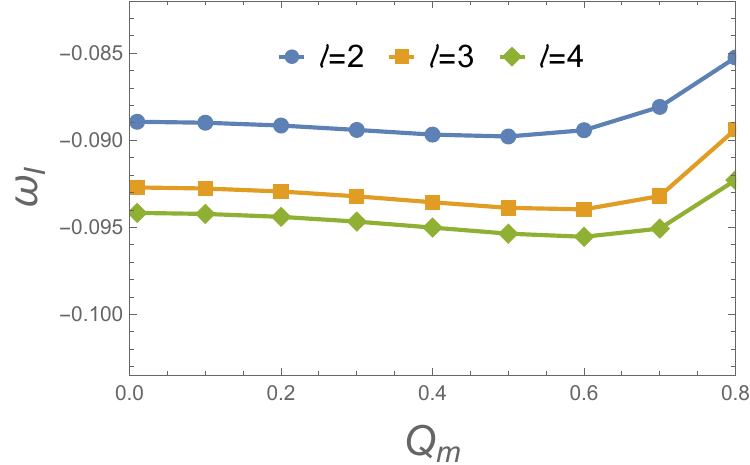}}
\caption{Variation of fundamental QNM frequencies  with respect to the magnetic charge $Q_m$ with $M=1$}\label{figQ1}
\end{figure}

Now, let's turn to discuss the influence of the magnetic charge $Q_m$ on the QNM frequencies of the lowest modes $(n=0)$ in gravitational field perturbation. These fundamental QNM frequencies with respect to different values $Q_m$ are plotted in Fig.\ref{figQ1}. For different branch solutions, these QNM frequencies exhibit distinct characteristics with increase of $Q_m$.
It is seen the real QNM frequencies for $\epsilon=0,1$ increase as $Q_m$ increases, and the damping rate or decay rate of perturbed field increases significantly with an increase of $Q_m$, see Figs. \ref{figQ11}-\ref{figQ16}. With regard to $\epsilon=-1$ in Figs. \ref{figQ13}-\ref{figQ14}), the real QNM frequencies also increase as $Q_m$ increases. However, the damping rate increases with the growth of $Q_m$, and then reach a maximum decay rate occurring at $Q_m=0.6$. As $Q_m$ surpasses this point, the damping rate begins to decrease.

\begin{figure}[H]
\centering
\subfigure[$ $]{\label{figeps11} 
\includegraphics[width=2.35in]{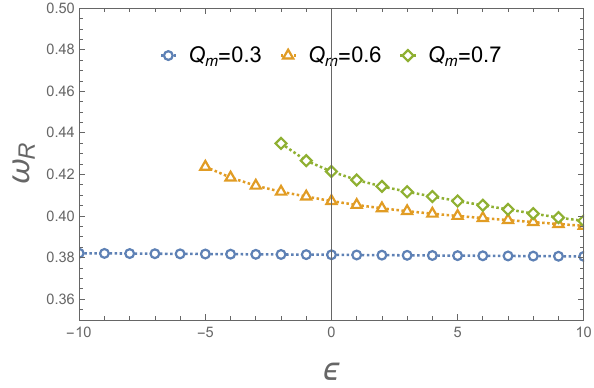}}
\subfigure[$ $]{\label{figeps12} 
\includegraphics[width=2.35in]{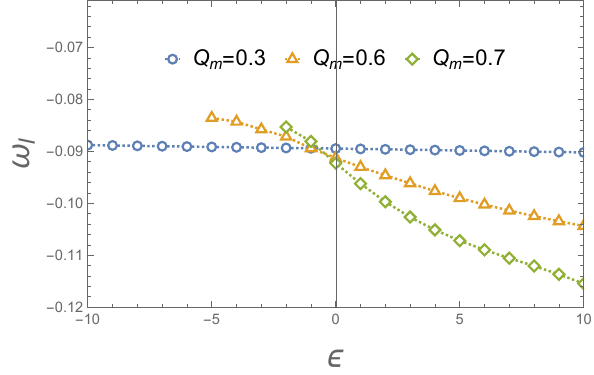}}
\caption{Variation of fundamental QNM frequencies with respect to $\epsilon$ with $M=1$ and $l=2$.}\label{figeps1}
\end{figure}

It's also interesting to investigate the effects of parameter $\epsilon$ on the QNM frequencies for gravitational perturbation. Fixed the magnetic charge $Q_m=0.3, 0.6$ and $0.7$, we display the variation of fundamental QNM frequencies and damping rate with different values of $\epsilon$ in Fig.\ref{figeps1}. As the parameter $\epsilon$ increases,  these QNM frequencies decrease for $Q_m=0.6$ and 0.7, while remain almost unchanged for $Q_m=0.3$. 

\section{Time domain integration}
\label{sec4}

To study QNM properties via gravitational wave propagation, we analyze the time-domain behavior by converting Eq. \eqref{perteq} into a wave equation, replacing the second-order term with $-\frac{d^2}{dt^2}$. This yields a generalized PDE for the gravitational field
\begin{eqnarray}
\left ( \frac{d^2}{dr_*}-\frac{d^2}{dt^2} -V(r)\right )\Psi(r,t)=0,
\end{eqnarray}
where the potential $V(r)$ is presented in \eq{oddpotv}.

Following the method in Refs.\cite{Zhu:2014sya,Yang:2021yoe,Fu:2022cul,Yang:2023lcm,Gogoi:2023fow}, we can
numerically solve above time-dependent wave-like equation. Here the finite difference method is applied for temporal integration, with a Gaussian wave serving as the initial spatial configuration. The radial coordinate is discretized based on the tortoise coordinate transformation
\begin{eqnarray}
\frac{dr (r_{*})}{dr_{*}} &=& \sqrt{A(r(r_*))B(r(r_*))}\Rightarrow \frac{r(r_{*j} + \Delta r_{*}) - r (r_{*j})}{\Delta r_{*}}\nonumber\\
& = &\frac{r_{j+1}-r_{j}}{\Delta r_{*}} = \sqrt{A(r_j)B(r_j)} \Rightarrow r_{j+1} = r_{j} + \Delta r_{*}\sqrt{A(r_j)B(r_j)} .
\end{eqnarray}

Then one can further discretize the effective potential into
$V(r(r_*)) = V(j\Delta r_*) = V_j$ and the field into $\Psi(r, t) =
\Psi(j\Delta r_*, i\Delta t) = \Psi_{j,i}$. Subsequently, the wave-like equation \cite{Zerilli:1970se} turns out to be a discretized equation
\begin{eqnarray}
-\frac{\Psi_{j,i+1} - 2\Psi_{j,i} + \Psi_{j,i-1}}{\Delta t^2} + \frac{\Psi_{j+1,i} - 2\Psi_{j,i} + \Psi_{j-1,i}}{\Delta r_*^2}
-V_j\Psi_{j,i} + \mathcal{O}(\Delta t^2) + \mathcal{O}(\Delta r_*^2) = 0,
\end{eqnarray}
from which one can isolate $\Psi_{j,i+1}$ after algebraic operations
\begin{eqnarray}
\Psi_{j,i+1} = \frac{\Delta t^2}{\Delta r_*^2}\Psi_{j+1,i} + \left(2 - 2\frac{\Delta t^2}{\Delta r_*^2} - \Delta t^2 V_j\right) \Psi_{j,i} + \Psi_{j-1,i} - \Psi_{j,i-1}.
\end{eqnarray}

The above equation is nothing but an iterative equation, which can be solved if one gives a Gaussian wave packet $\Psi_{j,0}$ as the initial perturbation. In our calculations,  setting the seed $r_{i=1}=r_h+10^{-12}$, after imposing the initial condition $\Psi_{j,i<0} = 0$, $\Psi_{j,0}=\exp[-\frac{(r_{j}-a)^2}{2b^2}]$. We choose the parameters $a = 40$ and $ b = 20$ in the Gaussian profile and setting $\frac{\Delta t}{\Delta r_*}=\frac{0.05}{0.1}=\frac{1}{2}$, this iterative equation provides us the evolution of the gravitational field perturbation in time profile.

\begin{figure}[H]
\centering
\subfigure[$\epsilon=1 $]{\label{figQ21} 
\includegraphics[width=2.in]{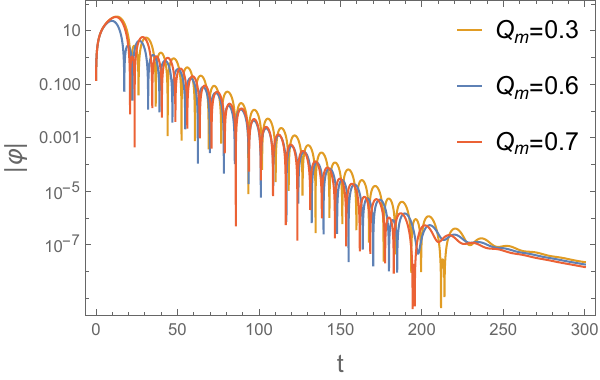}}
\subfigure[$\epsilon=0 $]{\label{figQ22} 
\includegraphics[width=2.in]{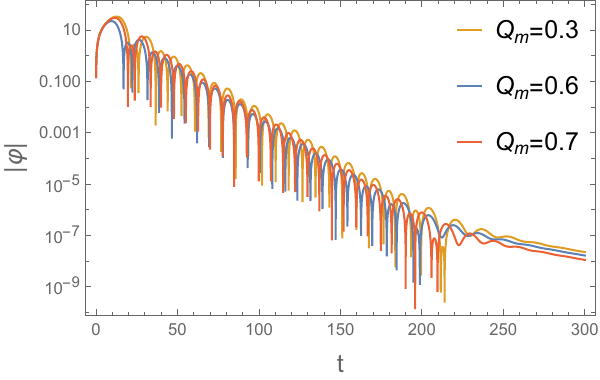}}
\subfigure[$\epsilon=-1 $]{\label{figQ23} 
\includegraphics[width=2.in]{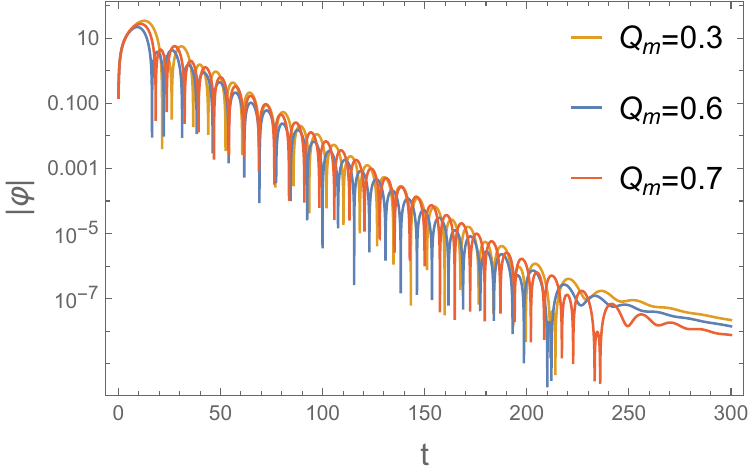}}
\caption{Time evolution for the perturbing gravitational field with $M=1$ and $l=2$.}\label{figQ2}
\end{figure}

The Fig.\ref{figQ2} presents the temporal profiles
for gravitational perturbation while varying the parameter $Q_m$. These profiles were obtained with an overtone number of $n=0$ and a multipole number of $l=2$ in both cases. For $\epsilon=0,1$ (see Figs.\ref{figQ21} and \ref{figQ22}), the large $Q_m$ makes the ringing stage of perturbation waveform more intensive and shorter, which corresponds to the larger $Re(\omega)$ and absolute value of $Im(\omega)$ (see Figs.\ref{figQ11}-\ref{figQ16}). 
In Fig. \ref{figQ23} for $\epsilon=-1$, the perturbation wave with $Q_m=0.6$ perform a shorter ringing stage, indicating larger values of absolute value of $Im(\omega)$. These agree well with the results in the frequency domain (Figs. \ref{figQ13} and \ref{figQ14}). Additionally, as a validation check, we employ the Prony method to fit the fundamental mode and the results are presented in Table. \ref{Table2}.

\begin{figure}[H]
\centering
 \subfigure[$Q_m=0.3$]{\label{figeps21} 
\includegraphics[width=2in]{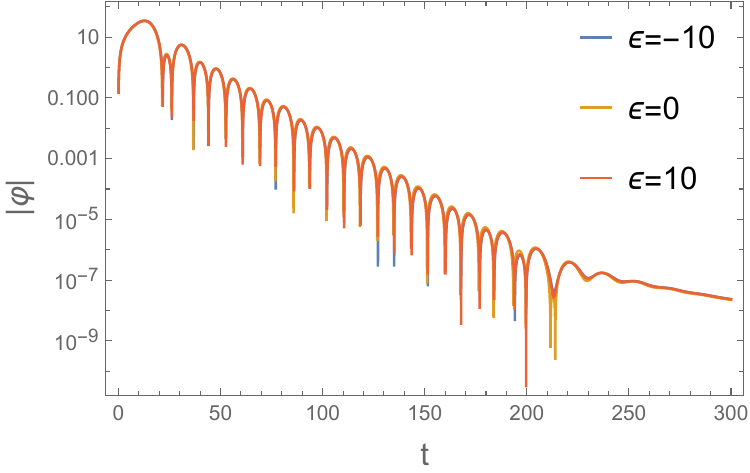}}
 \subfigure[$Q_m=0.6$]{\label{figeps22} 
\includegraphics[width=2.in]{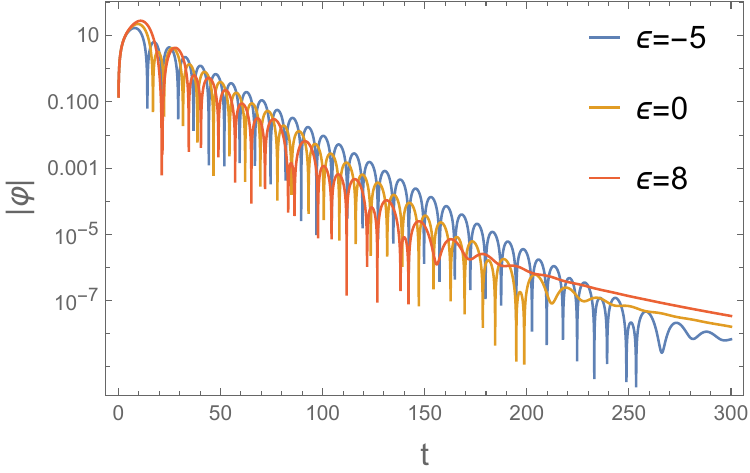}}
 \subfigure[$Q_m=0.7$]{\label{figeps23} 
\includegraphics[width=2.in]{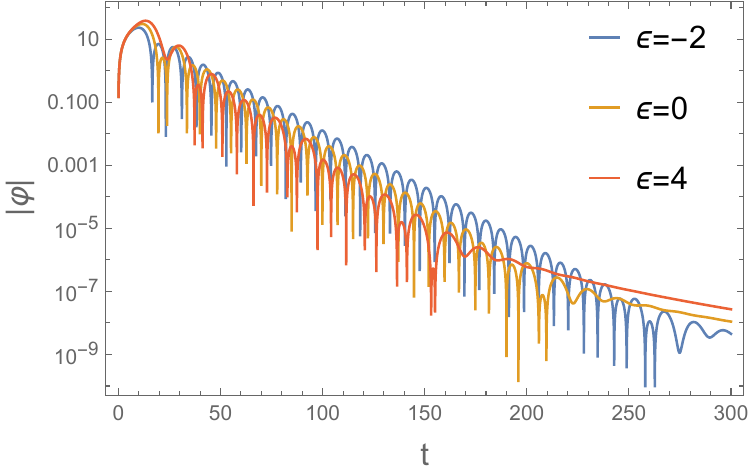}}
\hfill
\caption{Time evolution for the perturbing gravitational field with $M=1$ and $l=2$.}\label{figepS2}
\end{figure}

In Fig. \ref{figepS2}, we show the time domain profiles for
gravitational perturbation with different values of $\epsilon$. For $Q_m=0.3$, the real and imaginary parts of these QNM frequencies remain almost unchanged as the parameter $\epsilon$ vary. This stands in agreement with the variation of QNM plots studied in Fig.~\ref{figeps1} and potential behavior in Fig.\ref{fig30}. For $Q_m=0.6$ and 0.7, the large $\epsilon$ makes the ringing stage of perturbation waveform more sparser and shorter, which corresponds to the smaller $Re(\omega)$ and larger absolute value of $Im(\omega)$ (see Fig.\ref{figeps1}). 
Moreover, the QNM frequencies with different $\epsilon$ are also obtained by using Prony method, see Table. \ref{Table3}.


\begin{table*}[h!]
\caption{Fundamental QNM frequencies for gravitational field perturbation with $M=1$ and $l=2$.}\label{Table2}
\begin{tabular}{|m{2em}|c|c|c|} \hline
 	$Q_m$ &$\epsilon=1$&$\epsilon=0$ & $\epsilon=-1$ \\ \hline
 0.3 &  $0.381192-0.090218i$ & $0.381057-0.089666i$ & $0.382067 - 0.090253 i$       \\  \hline
  0.6 &  $0.405460-0.092283i$& $0.407387-0.090434i$ &$0.409135 - 0.089114 i$    \\ \hline
   0.7 &  $0.417572-0.096208i$&  $0.421273-0.092543i$ &$0.425926 - 0.087750  i$  \\ \hline
\end{tabular}
\end{table*}

\begin{table*}[h!]
\caption{Fundamental QNM frequencies for gravitational field perturbation with $M=1$ and $l=2$.}\label{Table3}
\begin{tabular}{|m{2em}|m{2em}|c|c|} \hline
$Q_m$ & $\epsilon$ & WKB method  &  Prony method \\ \hline
\multirow{3}{*}{0.3}& -10 &  $0.382084 - 0.0887889  i$ & $0.381169 - 0.0889776 i$ \\  
&  0 &  $0.381338 - 0.0894572 i $& $0.381057 - 0.0896657 i$   \\ 
 &  10 &  $0.380574 - 0.0902011 i$&  $0.380790 - 0.0895344 i$ \\ \hline
 \multirow{3}{*}{0.6}&  -5 &$0.423692 - 0.0835524 i$  & $0.423566 - 0.0835848 i$ \\  
&  0  &  $0.407187 - 0.0912665 i$ & $0.407387 - 0.0904337 i$  \\  
& 8  &  $0.397065 - 0.102459 i$  &  $0.396758 - 0.1022570 i$  \\  \hline
 \multirow{3}{*}{0.7}&  -2 &$0.434931 - 0.0853049 i$  & $0.434850 - 0.0856028 i$ \\ 
& 0  & $0.421409 - 0.0922722 i$ & $0.421495 - 0.0923237 i$ \\ 
&  4  &$0.409369 - 0.105113 i$  &$0.409109 - 0.1047280 i$\\ \hline
\end{tabular}
\end{table*}

\section{The Greybody factors}
\label{sec5}

In this section, we outline the application of the WKB method to the analysis of the greybody factor, a quantity that provides valuable insight into the transmission properties of the effective potential governing wave propagation in the given spacetime background. We begin by considering the wave equation under boundary conditions that allow for incoming waves originating from spatial infinity. Owing to the symmetry inherent in the scattering process, this configuration is mathematically equivalent to examining the scattering of a wave emanating from the black hole horizon. The appropriate boundary conditions for this scattering scenario are given by
\begin{eqnarray}
\psi&=&T(\omega)e^{-i\omega r_*}, \quad r_* \rightarrow -\infty,\\
\psi&=&e^{-i\omega r_*} + R(\omega)e^{i\omega r_*}, \quad r_* \rightarrow +\infty,
\end{eqnarray}
where $T$ is the transmission coefficient and $R$ is the reflection coefficient.
 
The square of the amplitude of the wave function at a particular point of spacetime determines the probability of finding it in the given point. The wave incoming towards a regular black hole is partially transmitted and partially reflected by the potential barrier. The greybody factor is defined as the probability of an outgoing wave reaching to infinity or an incoming wave absorbed by the black hole. Therefore, $|T(\omega)|^2$ is called the greybody factor, and $R(\omega)$ and $T(\omega)$ should satisfy the following relation
\begin{eqnarray}
|R(\omega)|^2+|T(\omega)|^2=1.
\end{eqnarray}
Using the 6th-order WKB method, the
reflection and transmission coefficients can be obtained
\begin{eqnarray}
&&|R(\omega)|^2 = \frac{1} {1 + e^{-2\pi i K(\omega)}} ,\nonumber\\
&&|T(\omega)|^2 =\frac{1} {1 + e^{2\pi i K(\omega)}}= 1 -|R(\omega)|^2,\label{TR}
\end{eqnarray}
where $K$ is a parameter which can be obtained by the WKB formula
\begin{eqnarray}
K= \frac{i\left( \omega^2 - V(r_0) \right)}{\sqrt{-2V''(r_0)}} + \sum_{i=2}^6 \Lambda_i.
\end{eqnarray}
For more detailed information, refer to reviews such as \cite{Konoplya:2019ppy,Konoplya:2011qq,Konoplya:2019hlu,Gogoi:2023fow} and the references therein.

\begin{figure}[H]
\centering
\subfigure[$\epsilon=-1$]{\label{figs11} 
\includegraphics[width=2in]{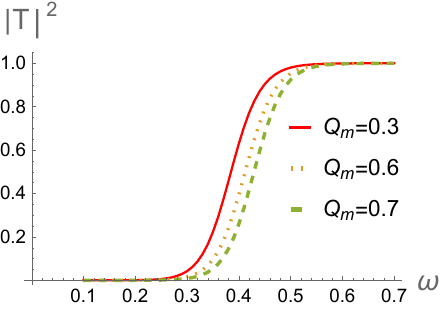}}
\hfill
\subfigure[$\epsilon=0$]{\label{figs12} 
\includegraphics[width=2in]{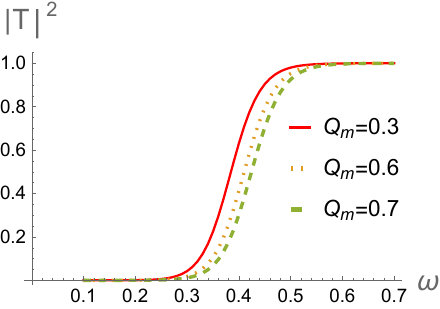}}
\hfill
\subfigure[$\epsilon=1$]{\label{figs13} 
\includegraphics[width=2in]{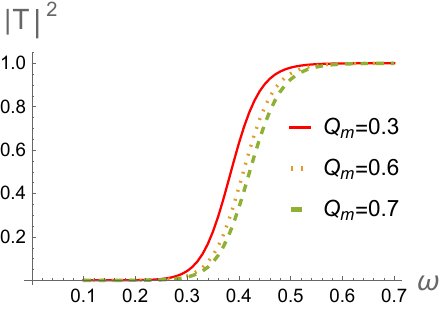}}
\hfill
\caption{The greybody factors for the gravitational field with $M=1$ and $l=2$.}\label{figs1}
\end{figure}

\begin{figure}[H]
\centering
\subfigure[$\epsilon=-1$]{\label{figs21} 
\includegraphics[width=2.in]{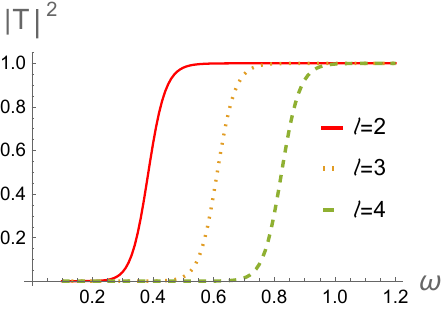}}
\hfill
\subfigure[$\epsilon=0$]{\label{figs22} 
\includegraphics[width=2.in]{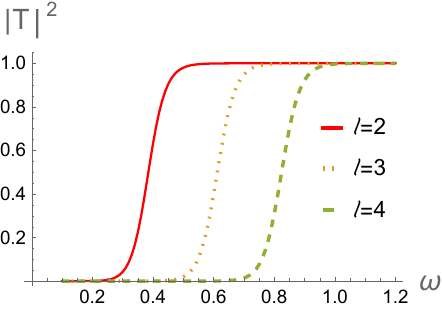}}
\hfill
\subfigure[$\epsilon=1$]{\label{figs23} 
\includegraphics[width=2.in]{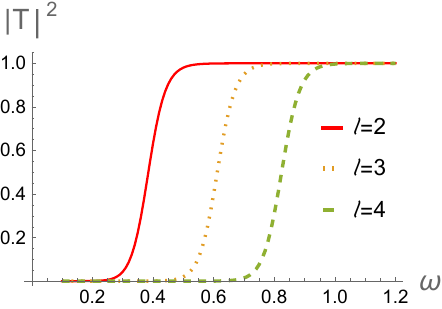}}
\hfill
\caption{The greybody factors for the gravitational field with $M=1$ and $l=2$.}\label{figs2}
\end{figure}

\begin{figure}[H]
\centering
\subfigure[$Q_m=0.3$]{\label{figs31} 
\includegraphics[width=2in]{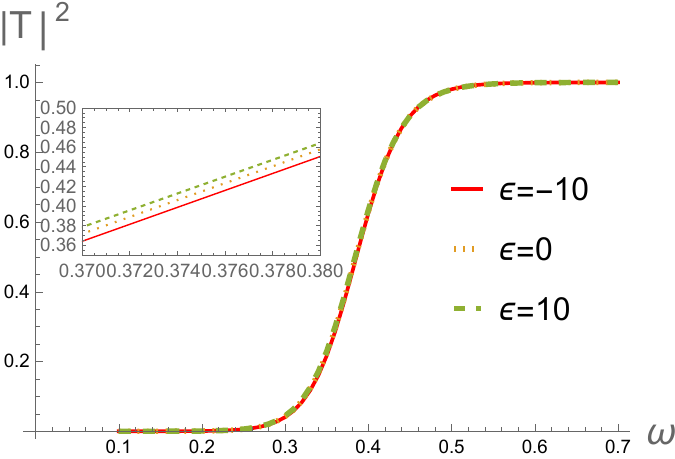}}
\hfill
\subfigure[$Q_m=0.6$]{\label{figs32} 
\includegraphics[width=2in]{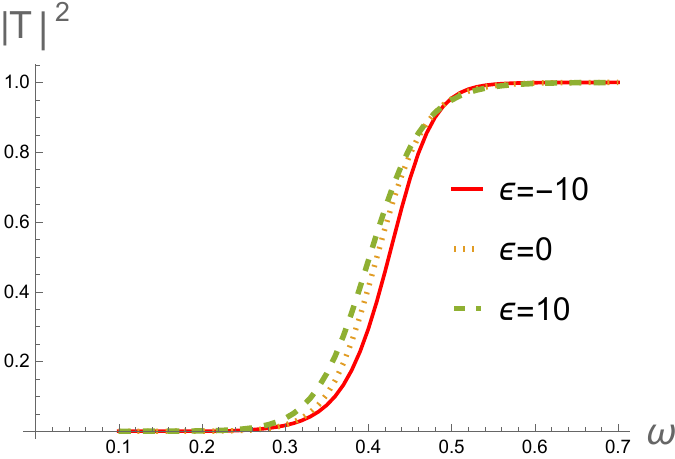}}
\hfill
\subfigure[$Q_m=0.7$]{\label{figs33} 
\includegraphics[width=2.in]{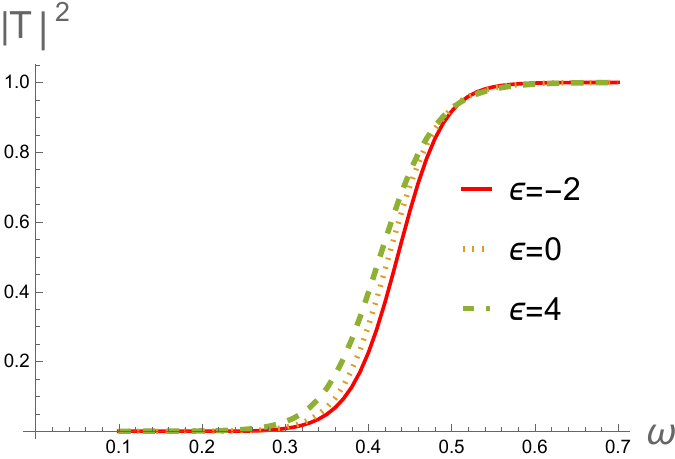}}
\hfill
\caption{The greybody factors for the gravitational field with $M=1$ and $l=2$.}\label{figs3}
\end{figure}

In Figs.~\ref{figs1}-\ref{figs3}, we show the behaviors of the greybody factors for gravitational perturbation under varying the magnetic charge $Q_m$, multipole number $l$ and parameter $\epsilon$. It's seen that the greybody factors exhibit noticeable decrease as $Q_m$ increases (see Fig.\ref{figs1}). This
suggests that when the magnetic charge $Q_m$ is reduced, the black hole becomes more effective at capturing and interacting with incoming matter or radiation. We also investigate how a change of multipole number $l$ in affects the corresponding behavior of the greybody factors (see Fig.\ref{figs2}). The greybody factors also gradually decrease as $l$ increases, which reveals that the greybody factors are higher for smaller values of $l$. With regard to difference $\epsilon$, the greybody factors gradually increase as $\epsilon$ increases (see Fig.\ref{figs3}). It indicates that black holes become less interactive with the surrounding radiation and allow more of perturbed field to escape. These are all consistent with the effective potentials in Figs.\ref{figpov1}-\ref{figpov3}.

\section{Conclusion and discussion}
\label{sec6}

In this study, we examined axial field perturbations of magnetically charged black holes within the framework of string-inspired Euler–Heisenberg theory. Under the consideration of axial gravitational perturbations decoupled from axial electromagnetic perturbations, we systematically analyze the effects of the magnetic charge $Q_m$, the parameter $\epsilon$, and the angular quantum number $l$ on the corresponding effective potentials associated with the axial gravitational perturbations.

Then, we adopted the AIM and Pad\'e-averaged 6th order WKB methods to compute QNMs in each scenario.
We considered how QNMs changes with the multipole moment $l$, magnetic charge $Q_m$ and parameter $\epsilon$.
For $\epsilon=0,1$, we found that the real part of QNM frequencies increase as $Q_m$ increases, and the damping rate or decay rate of gravitational waves increases significantly with increase of $Q_m$. With regard to $\epsilon=-1$,
the real part of QNMs frequencies also increase as $Q_m$ increases. However, the damping rate increase with the growth of $Q_m$, and then reach a maximum decay rate occurring at $Q_m=0.6$. As $Q_m$ surpasses this point, the damping rate slowly begins to decrease. On the other hand,  we found that these QNM frequencies decrease for $Q_m=0.6$ and 0.7, while remain almost unchanged for $Q_m=0.3$ as the parameter $\epsilon$ increases. 
These observations in frequency domains agree well with the results we obtained in the time domains.

Using the sixth-order WKB approximation, we have also computed the greybody factor associated with the gravitational field perturbations. The results indicate that the magnetic charge $Q_m$ has a suppressing effect on the tunnelling probability-specifically, a smaller fraction of the perturbed field is able to penetrate the effective potential barrier as $Q_m$ increases. In contrast, varying the parameter $\epsilon$ gives rise to an opposite trend. These observations are in agreement with the qualitative behavior of the corresponding effective potential profiles.


Notice that Ref.\cite{Daghigh:2021psm} recovered the axial metric perturbation decouples from axial electromagnetic perturbation for magnetic black holes. Some papers \cite{Nomura:2020tpc,Meng:2022oxg,Daghigh:2021psm} further discussed the combination: axial gravitational perturbation coupled with polar electromagnetic perturbation or polar gravitational perturbation coupled with axial electromagnetic perturbation for the magnetic regular black holes.
We will consider these combinations for the magnetically charged black holes in the future works.

 \vspace{1cm}
{\bf Acknowledgments}
 \vspace{0.5cm}

We appreciate Guoyang Fu and Zhen-Hao Yang for helpful
discussions. This research is supported by the National Natural Science Foundation of China (NNSFC) (Grant No.12365009) and Natural Science Basic Research Program of Shaanxi (Program No.2023-JC-QN-0053). 

\appendix
\section{Axial electromagnetic perturbation}

In this appendix, we will show the perturbed equation for axial electromagnetic perturbation. Substituting the perturbed metric and vector potential (Eqs.\eqref{pertmetric}, \eqref{eqodd} and \eqref{vectorpert}) into the Maxwell equation \eqref{eqMaxwell}, we obtain
\begin{eqnarray}
E^{(A)}_{t}&=&h_0(r) \Big[4 Q_m^2 e^{2 \phi (r)} (\alpha +2 \beta +3 \beta  \cosh (2 \phi (r)))+r^4\Big],\label{app1}\\
E^{(A)}_{r}&=&h_1(r) \Big[4 Q_m^2 e^{2 \phi (r)} (\alpha +2 \beta +3 \beta  \cosh (2 \phi (r)))+r^4\Big],\label{app2}\\
E^{(A)}_{\theta}&=&4 Q_m^2 e^{2 \phi (r)} \Big[\alpha +2 \beta +3 \beta  \cosh (2 \phi (r))\Big]\Big[r B(r) h_1(r) A'(r)+A(r) \left(r h_1(r) B'(r)\right.\nonumber\\
&&\left.+2 B(r) \left(r h_1'(r)-6h_1(r)\right)\right)+2 i r \omega h_0(r)\Big]+r^5 B(r) A'(r) h_1(r)\nonumber\\
&&+24 \beta  Q^2 r A(r) B(r) e^{4 \phi (r)} \phi '(r) h_1(r)+2 i r^5
   \omega h_0(r)\nonumber\\
&&+r A(r) \Big[r^4 B'(r) h_1(r)+2 B(r) \left(r^4 h_1'(r)-2
   h_1(r) \left(\left(6 \beta  Q^2+r^4\right) \phi'(r)+r^3\right)\right)\Big],\label{app3}\\
E^{(A)}_{\varphi}&=&\Big[\omega ^2-\frac{3
l(l+1) A(r)}{r^2}+\frac{2 (l+1) l r^2 A(r)}{4 Q_m^2 e^{2 \phi (r)} (\alpha +2 \beta+3\beta\cosh (2 \phi (r)))+r^4}\Big]u_3(r)\nonumber\\
&&+\Big[\frac{4 A(r) B(r) \left(\phi '(r) \left(-6 \beta Q_m^2-2
Q_m^2 (\alpha +2 \beta ) e^{2 \phi (r)}-r^4\right)+r^3\right)}{4 Q_m^2 e^{2 \phi (r)} (\alpha +2 \beta +3 \beta  \cosh (2
\phi (r)))+r^4}\nonumber\\
&&+2 A(r) B(r) \phi '(r)-\frac{4 A(r) B(r)}{r}+\frac{1}{2} B(r) A'(r)+\frac{1}{2} A(r) B'(r)\Big]u_3'(r) \nonumber\\
&&+A(r) B(r) u_3''(r).\label{app4}
\end{eqnarray}
From these equations, it's seen that the gravitational perturbation functions $h_0(r)$ and $h_1(r)$ must vanish and there only exist single perturbation equation for perturbed electromagnetic field. 

In order to cast this master equation \eqref{app4} into the standard Schr\"odinger form, we further define the function $u_3(r)$ with
\begin{eqnarray}
u_3(r)=C_1(r)*\Psi_A(r).
\end{eqnarray}
We assume the function $C_1(r)$ taking following form
\begin{eqnarray}
C_1(r)=\frac{r^2 e^{\phi (r)}}{\sqrt{2 Q^2 \left(3 \beta +2 (\alpha +2 \beta ) e^{2 \phi (r)}+3 \beta  e^{4 \phi (r)}\right)+r^4}},
\end{eqnarray}
and then obtain the final perturbed equation
\begin{eqnarray}
 \frac{d^2\Psi_A(r_*)}{d r^2_{*}}+\Big[\omega^2-V_A(r)\Big]\Psi_A(r_*)=0,
\end{eqnarray}
where $r_*$ represents the tortoise coordinate with
$d{r_*} = \frac{1}{\sqrt{AB}}dr$
and $V_A(r)$ denotes the effective potential for axial electromagnetic perturbation. Due to $V_A(r)$ cumbersome expression, its specific form is not provided here.

\end{document}